\documentclass[pdflatex,letterpaper,11pt]{article}
\usepackage[utf8]{inputenc}
\usepackage{fullpage}
\usepackage{authblk}
\usepackage{graphics}
\usepackage[pdftex]{graphicx}
\usepackage{amsfonts}
\usepackage{microtype}
\usepackage{wrapfig}
\usepackage{amsmath}
\usepackage{amssymb}
\usepackage{booktabs} 
\usepackage{cite}
\usepackage{xcolor}
\usepackage[format=plain,labelformat={simple},labelsep={period},labelfont={bf}]{caption}
\usepackage{microtype}
\usepackage[T1]{fontenc}
\usepackage{lmodern}
\usepackage{hyperref}

\newcommand{\Xcomment}[1]{}

\def\comment#1{}

\def\withcomments{
\newcounter{mycommentcounter}
\def\comment##1{\refstepcounter{mycommentcounter}
\ifhmode
\unskip
{\dimen1=\baselineskip \divide\dimen1 by 2 
\raise\dimen1\llap{\tiny\bfseries \textcolor{red}{-\themycommentcounter-}}}\fi
\marginpar[{\renewcommand{\baselinestretch}{0.8}
\hspace*{3em}\begin{minipage}{5em}\footnotesize [\themycommentcounter] \raggedright ##1\end{minipage}}]{\renewcommand{\baselinestretch}{0.8}
\begin{minipage}{5em}\footnotesize [\themycommentcounter]: \raggedright ##1\end{minipage}}}
}

\newcommand{\dist}{\mathit{dist}}
\newcommand{\cost}{\mathit{cost}}
\newcommand{\redcost}{\bar{c}}

\newcommand{\mst}{\mbox{MST}}

\newcommand{\vorbase}{\mathit{vb}}
\newcommand{\vordist}{\mathit{vd}}
\newcommand{\vorparent}{\mathit{vp}}

\newcommand{\instance}[1]{\textsf{#1}}
\newcommand{\class}[1]{\instance{#1}}
\newcommand{\sseries}[1]{\instance{#1}}
\newcommand{\series}[1]{\instance{#1}}
\newcommand{\tabhead}[1]{\textsc{#1}}
\newcommand{\optsol}{\ensuremath{\mbox{\textsc{opt}}}}

\title{\textbf{A Robust and Scalable Algorithm for the\\Steiner Problem in Graphs}\thanks{This work was presented at the 11th DIMACS Implementation Challenge in Collaboration with ICERM: Steiner Tree Problems~\cite{dimacs11}. It was partly done while R.\,Werneck was at Microsoft Research Silicon Valley.}}
\author[1]{Thomas Pajor}
\author[2]{Eduardo Uchoa}
\author[3]{Renato F.\ Werneck}
\affil[1]{Microsoft Research, USA, \href{mailto:tpajor@microsoft.com}{tpajor@microsoft.com}}
\affil[2]{Universidade Federal Fluminense, Brazil, \href{mailto:uchoa@producao.uff.br}{uchoa@producao.uff.br}}
\affil[3]{San Francisco, USA, \href{mailto:rwerneck@cs.princeton.edu}{rwerneck@cs.princeton.edu}}

\begin{document}
\maketitle

\begin{abstract}
We present an effective heuristic for the Steiner Problem in Graphs. Its main elements are a multistart algorithm coupled with aggressive combination of elite solutions, both leveraging recently-proposed fast local searches. We also propose a fast implementation of a well-known dual ascent algorithm that not only makes our heuristics more robust~(by quickly dealing with easier cases), but can also be used as a building block of an exact~(branch-and-bound) algorithm that is quite effective for some inputs. On all graph classes we consider, our heuristic is competitive with~(and sometimes more effective than) any previous approach with similar running times. It is also scalable: with long runs, we could improve or match the best published results for most open instances in the literature.
\end{abstract}

\newcommand{\coverage}[0]{\ensuremath{\mathcal{C}}}

\newcommand{\selfnote}[1]{\footnote{\sl #1}}

\section{Introduction}
\label{sec:intro}

Given an edge-weighted, undirected graph~$G = (V,E)$ and a set~$T \subseteq V$ of terminals, the Steiner Problem in Graphs~(SPG) is that of finding a minimum-cost tree that contains all vertices in~$T$. This has application in many areas, including computational biology, networking, and circuit design~\cite{CD02}. Unfortunately, it is NP-hard not only to find an optimal solution~\cite{Kar72}, but also to approximate it within a factor of 96/95~\cite{CC02}. The best known approximation ratio is 1.39~\cite{BGRS10}~(see~\cite{RZ05} for a 1.55 approximation). Given its practical importance, there is a wealth of exact algorithms~\cite{DV99,KM98,PUW01,Pol03,PV01b,RUW02,UPR02,Vah03,LLLPR14,HSV14} and heuristics~\cite{BR01,DV97,DV99,PRUW01,RS00,RUW02,BBZ12,UW12} to deal with real-world instances. State-of-the-art algorithms use a diverse toolkit that includes linear relaxations, branch-and-bound, reduction tests~(preprocessing), and primal and dual heuristics.

Our goal in this paper is to develop an algorithm that is, above all, \emph{robust}. For any input instance, regardless of its characteristics, we want to quickly produce a good solution. Moreover, the algorithm should \emph{scale} well: when given more time to run, it should produce better solutions.

Our basic algorithm follows the principles of a heuristic proposed by Ribeiro et al.~\cite{RUW02}: it is a multistart algorithm with recombination~(a genetic component), using perturbation for randomization. Under the hood, however, we introduce significant improvements that lead to much better results.

First, we leverage fast local search algorithms recently proposed by Uchoa and Werneck~\cite{UW12}, which are asymptotically faster~(in theory and practice) than previous approaches. Second, we propose a \emph{cascaded combination} strategy, which combines each fresh~(newly-created) solution with multiple entries from a pool of elite solutions, leading to much quicker convergence. Third, we counterbalance this intensification strategy with a series of diversification measures~(including more careful perturbation and replacement policies in the pool) in order to explore the search space more comprehensively.

As a result, long runs of our algorithm can match or even improve the best published solutions~(at the time of writing) for several open instances in the literature. For easier inputs, our basic algorithm still finds very good results, but for some graph classes it can be slower than alternative approaches that rely heavily on preprocessing and small duality gaps~\cite{Pol03,Vah03}.

To make our overall approach more robust, we include some basic preprocessing routines. Moreover, we propose a \emph{Guarded Multistart} algorithm, which runs a branch-and-bound routine \emph{at the same time} as our basic~(primal-only) algorithm. For easy instances, these two threads can help each other, often leading to drastic reductions in total CPU time. To compute lower bounds, we propose a novel and efficient implementation of a well-known combinatorial dual ascent algorithm due to Wong~\cite{Won84}. For several hard instances, our branch-and-bound routine finds provably optimal solutions faster than any published algorithm.

Even with these optimizations, there are important graph classes~(such as some VLSI instances) in which our method is not as effective as other approaches, notably those based on advanced reduction techniques and linear programming~\cite{Vah03,Pol03} or on dynamic programming~\cite{HSV14}~(when the number of terminals is small). Even in such cases, however, the solutions found by our approach are not much worse, confirming its robustness. Overall, our algorithm provides a reliable, general-purpose solution for the Steiner Problem in Graphs.

This paper is organized as follows. Section~\ref{sec:meta} explains our multistart algorithm. Section~\ref{sec:lower} discusses our lower-bounding techniques, including branch-and-bound. Section~\ref{sec:robust} shows how preprocessing and lower-bounding make our basic algorithm more robust. Section~\ref{sec:experiments} has experiments, and we conclude in Section~\ref{sec:conclusion}.

\paragraph{Notation.} The input to the Steiner Problem in Graphs is an undirected graph~$G = (V,E)$ and a set~$T \subseteq V$ of \emph{terminals}. Each edge~$e = (v,w)$ has an associated nonnegative \emph{cost}~(\emph{length}) denoted by~$\cost(e)$ or~$\cost(v,w)$. A \emph{solution}~$S = (V_S, E_S)$ is a tree with~$T \subseteq V_S \subseteq V$ and~$E_S \subseteq E$; its cost is the sum of the costs of its edges. Our goal is to find a solution of minimum cost.

\section{Basic Algorithm}
\label{sec:meta}

Our basic algorithm follows the multistart approach and runs in~$M$ iterations~(where~$M$ is an input parameter). Each iteration generates a new solution from scratch using a constructive algorithm~(with randomization), followed by local search. We also maintain a pool of elite solutions with the best and other good solutions found so far. The main feature of our algorithm is a \emph{cascaded combination} strategy, which aggressively combines a new solution with several existing ones. This finds very good solutions soon, but has a very strong intensification effect. To counterbalance it, we look for diversification in other aspects of the algorithm. The outline of our algorithm is as follows:

\begin{enumerate}
\item Create an empty pool~$P$ of elite solutions with capacity~$\lceil{\sqrt{M/2}}\rceil$.
\item Repeat for~$M$ iterations:
\begin{enumerate}
\item\label{stp:newsol} Generate a new solution~$S$ using a constructive algorithm, local search, and randomization.
\item Generate a solution~$S'$ by combining~$S$ with solutions in the pool.
\item Try to add~$S$ and then~$S'$ to the pool~$P$.
\end{enumerate}
\item Return the best solution in the pool~$P$.
\end{enumerate}

The remainder of this section describes details omitted from this outline. Section~\ref{sec:localsearch} explains the local search routines, Section~\ref{sec:newsol} describes how fresh solutions are generated, Section~\ref{sec:combination} deals with the cascaded combination algorithm, and Section~\ref{sec:pool} addresses the insertion and eviction policies for the pool.

\subsection{Local Search}
\label{sec:localsearch}

A local search algorithm tries to improve an existing solution~$S$ by examining a \emph{neighborhood}~${\cal N}(S)$ of~$S$, a set of solutions obtainable from~$S$ by performing a restricted set of operations. \emph{Evaluating}~${\cal N}(S)$ consists of either finding an improving solution~$S'$~(i.e., one with~$cost(S') < cost(S)$) or proving that no such~$S'$ exists in~${\cal N}(S)$. A local search heuristic repeatedly replaces the current solution by an improving neighbor until it reaches a \emph{local minimum}~(or \emph{local optimum}). Uchoa and Werneck~\cite{UW12} present algorithms to evaluate in~$O(|E| \log |V|)$ time four natural~(and well-studied) neighborhoods: Steiner-vertex insertion, Steiner-vertex elimination, key-path exchange, and key-vertex elimination.

The first two use the representation of a solution~$S = (V_S,E_S)$ in terms of its set~$V_S \setminus T$ of \emph{Steiner vertices}. The minimum spanning tree~(MST) of the subgraph of~$G$ induced by~$V_S$~(which we denote by~$\mst(G[V_S])$) costs no more than~$S$. In particular, if~$S$ is optimal, so is~$\mst(G[V_S])$. Uchoa and Werneck use dynamic graph techniques to efficiently evaluate neighborhoods defined by the insertion or removal of a single Steiner vertex~\cite{Min90,OG91,SP75,Vos92}. Although such neighborhoods had been used in metaheuristics before~\cite{BR01,RS00,RUW02}, they were much slower:~$O(|V|^2)$ time for insertions and~$O(|E| |V|)$ time for removals

The other two neighborhoods represent a solution~$S$ in terms of its \emph{key vertices}~$K_S$, which are Steiner vertices with degree at least three in~$S$. If~$S$ is optimal, it costs the same as the MST of its \emph{distance network} restricted to~$K_S \cup T$~(the complete graph on~$|K_S \cup T|$ vertices whose edge lengths reflect shortest paths in~$G$). Uchoa and Werneck show that the neighborhood corresponding to the elimination of a single key vertex can be evaluated in~$O(|E| \log |V|)$ time. They prove the same bound for the \emph{key-path exchange} local search~\cite{Dow91,DV97,VSA96}, which attempts to replace an existing key path~(linking two vertices of~$K_S \cup T$ in~$S$) by a shorter path between the components it connects. Both implementations improve on previous time bounds by a factor of~$O(|T|)$.

In this paper, we mostly take the local searches as black boxes, but use the fact that these implementations work in \emph{passes}. If there is an improving move in the neighborhood, a pass is guaranteed to find one in~$O(|E| \log |V|)$ time. To accelerate convergence, the algorithms may perform multiple independent moves in the same pass~(within the same time bound). In practice, there are almost always fewer than 10 passes, with most of the improvements achieved early on~\cite{UW12}.

We follow Uchoa and Werneck and use what they call the \textsc{vq} local search within our algorithm. It alternates between a pass that evaluates Steiner-vertex insertion~(\textsc{v}) and a pass that evaluates~(simultaneously) both key-vertex removal and key-path exchange~(\textsc{q}). Since in practice Steiner vertex removal~(\textsc{u}) is rarely better than key-vertex removal, our main algorithm does not use it.

\subsection{Generating New Solutions}
\label{sec:newsol}

New solutions are generated by a constructive algorithm followed by local search, using randomization. Instead of making our algorithms~(constructive and local search) randomized, we follow Ribeiro et al.~\cite{RUW02} and apply perturbations to the edge costs instead, preserving the running time guarantees of all algorithms.

\paragraph{Using the perturbation.} To build a constructive solution, we apply a random perturbation to the edge costs~(details will be given later), then run a near-linear time~(in practice) implementation~\cite{PW02} of the shortest-path heuristic~(SPH)~\cite{TM80}. Starting from a random root vertex, SPH greedily adds to the solution the entire shortest path to the terminal that is closest~(on the perturbed graph) to previously picked vertices.

Ribeiro et al.~\cite{RUW02} suggest applying local search to the constructive solution, but using the original~(unperturbed) costs during local search. Since the constructive solution can be quite far from the local optimum, however, the effects of the perturbation tend to disappear quite soon, hurting diversification.

We propose an alternative approach, which leverages the fact that our local searches work in passes. We start the local search on the perturbed instance and, after each pass it makes, we \emph{dampen} the perturbation, bringing all costs closer to their original values. For each edge~$e$ with original~(unperturbed) cost~$\cost(e)$, let~$\cost_i(e)$ be its~(perturbed) cost at the end of pass~$i$. For pass~$i+1$, we set~$\cost_{i+1}(e) = \alpha \cost_i(e) + (1 - \alpha) \cost(e)$, where~$0 < \alpha < 1$ is a \emph{decay factor}~(we use~$\alpha = 0.5$). This approach makes better use of the guidance provided by the perturbation, thus increasing diversification. For efficiency, after three passes with perturbation, we restore the original~(unperturbed) costs and run the local search until a local optimum is reached.

\paragraph{Computing the perturbation.} We now return to the issue of how initial perturbations are computed. Ribeiro et al.\ propose a simple \emph{edge-based} approach, in which the cost of each edge is multiplied by a random factor~(chosen between 1.0 and 1.2 in their case). This is reasonably effective, but has a potential drawback: because edges are independent, the perturbations applied to each incident edge to any particular vertex tend to cancel out one another. We thus propose a \emph{vertex-based} perturbation, which associates an independent random factor to each vertex in the graph. The perturbed cost of an edge~$(u,v)$ is then the original cost multiplied by the average factors of its two endpoints~($u$ and~$v$).

To enhance diversification, the choice of parameters that control the perturbation itself is randomized. Each iteration chooses either edge-based or vertex-based perturbation with equal probability. It then picks a maximum perturbation~$Q$ uniformly at random in the range~$[1.25,2.00]$. Finally, it defines the actual perturbation factors: for each element~(vertex or edge), it sets the factor to~$1 + \rho Q$, where~$\rho$ is generated~(for each element) uniformly at random in~$[0,1]$.

To achieve further diversification, we actually use a slightly non-uniform distribution parameterized by a small threshold~$\tau = (\log_2 n) / n$. If the random number~$\rho$ is at least~$\tau$~(as is usually the case), we use the formula above. Otherwise, we use a perturbation factor of~$\rho/\tau$. Note that this factor is between 0.00 and 1.00, while the standard factor is always between 1.25 and 2.00. This means that a small fraction of the elements can become significantly cheaper than others, and are thus more likely to appear in the solution. This allows us to test key-vertices that our standard local searches would not normally consider~(recall that we do not have a fast local search based on key-vertex insertion).

We stress that the algorithm already works reasonably well with the standard edge-based perturbation proposed by Ribeiro et al.~\cite{RUW02}; although we observed some improvement with the vertex-based perturbation~(and the non-uniform distribution), the effects were relatively minor.

\subsection{Cascaded Combination}
\label{sec:combination}

The cascaded combination algorithm takes as input an initial solution~$S_0$, the pool of elite solutions, and the \emph{maximum number of allowed failures}, denoted by~$\phi$~(we use~$\phi=3$). The procedure combines~$S_0$ with elements in the pool, generating a~(potentially better) solution~$S^*$.

The basic building block of this procedure is the \emph{randomized merge} operation~\cite{RUW02}, which takes as input two solutions~($S_a$ and~$S_b$) and produces a third~(potentially cheaper) one. It does so by first generating a \emph{perturbed graph}~$G'$ from~$G$ by perturbing each edge cost depending on which of the original solutions~($S_a$ and/or~$S_b$) it appears in. If an edge appears in both solutions, it keeps its original cost. If it appears in none of the solutions, its cost is multiplied by 1000. If it appears in exactly one solution~($S_a$ or~$S_b$), its cost is multiplied by a random number between 100 and 500. We run the SPH heuristic on the resulting instance. We then remove all perturbations and apply local search to the combined solution, producing~$S_c$, the result of the perturbed combination.

The cascaded combination procedure maintains an incumbent solution~$S^*$, originally set to~$S_0$. In each step, it performs a randomized merge of~$S^*$ and a solution~$S'$ picked uniformly at random from the pool. Let~$S''$ be the resulting solution. If~$\cost(S'') < \cost(S^*)$, we make~$S''$ the new incumbent~(i.e., we set~$S^* \leftarrow S''$). Otherwise, we say that the randomized merged \emph{failed} and keep~$S^*$ as the incumbent. When the number of failures reaches~$\phi$, the cascaded combination algorithm stops and returns~$S^*$.

Note that the resulting solution~$S^*$ may have elements from several other solutions in the pool. This makes it a powerful intensification agent, helping achieve good solutions quite quickly. That said, the first few solutions added to the pool will have a disproportionate influence on all others, potentially confining the multistart algorithm to a very restricted region of the search space. This is why we prioritize diversification elsewhere in the algorithm.

On average, each multistart iteration touches a constant number of solutions in the pool. We set the capacity of the elite pool to~$\Theta(\sqrt{M})$ to ensure that most pairs of elite solutions are~(indirectly) combined with one another at some point during the algorithm. We set the precise capacity to~$\lceil \sqrt{M/2} \rceil$, but the algorithm is not too sensitive to this constant; results were not much different with~$\lceil \sqrt{M} \rceil$ or~$\lceil \sqrt{M/4} \rceil$.

\subsection{Pool Management}
\label{sec:pool}

We now address the insertion and eviction policies for the pool of elite solutions. When our algorithm attempts to add a solution~$S$ to the pool, we must consider three simple cases and a nontrivial one. First, if~$S$ is identical to a solution already in the pool, it is not added. Second, if the pool is not full and~$S$ is not identical to any solution,~$S$ is simply added. Third, if the pool is full and~$S$ is not better than any solution in the pool,~$S$ is not added.

The nontrivial case happens when the pool is full,~$S$ is different from all solutions in the pool, and~$S$ is better than the worse current solution. In this case,~$S$ replaces a solution that is at least as bad as~$S$, with~(randomized) preference for solutions that are similar to~$S$~(based on the symmetric difference between their edge sets). This technique has been shown to increase diversification for other problems~\cite{RW04}.

\subsection{Discussion}

As Section~\ref{sec:experiments} will show, our algorithm significantly outperforms the multistart approach by Ribeiro et al.~\cite{RUW02}. Although there are many differences between the algorithms, the main reason for our good performance is that we leverage much faster implementations of the underlying local searches~\cite{UW12}. Not only can we run more iterations within the same time limit, but we can also do more in each iteration---with cascaded combinations.

Although relevant, other aspects of our algorithm have less effect than these two factors~(fast local searches and cascaded combinations). The parameters we report~(and there are many) are the ones we ended up using in the final version of our code. In most cases, however, the algorithm is not very sensitive to the exact value of these parameters, as long as they are reasonable. This includes the size of the pool, the type of perturbation, and the criteria for selecting elements from the pool.

\section{Lower Bounds}
\label{sec:lower}

We now turn our attention to finding lower bounds. We use an efficient implementation of an existing greedy algorithm~(due to Wong~\cite{Won84}) associated with a powerful linear programming formulation. We first define the formulation and an abstract version of the algorithm, and then discuss our implementation.

\subsection{Formulation and Dual Ascent}
\label{sec:wongcut}

We use the dual of the well-known directed cut formulation for the Steiner problem in graphs~\cite{Won84}. It takes a terminal~$r \in T$ as the root. A set~$W \subset V$ is a \emph{Steiner cut} if~$W$ does not contain the root but contains at least one terminal. Let~$\delta^-(W)$ be the set consisting of all arcs~$(u,v)$ such that~$u \not\in W$ and~$v \in W$.

The dual formulation associates a nonnegative variable~$\pi_W$ with each Steiner cut~$W$. The set~$\cal W$ of all Steiner cuts has exponential size. Given a dual solution~$\pi$, the \emph{reduced cost}~$\redcost(a)$ of an arc~$a$ is defined as~$\cost(a) - \sum_{W \in {\cal W}: a \in \delta^-(W)} \pi_W$. The dual formulation maximizes the sum of all~$\pi_W$ variables, subject to all reduced costs being nonnegative. An arc whose reduced cost is zero is said to be \emph{saturated}.

The basic building block of exact algorithm is a \emph{dual ascent} routine proposed by Wong~\cite{Won84}, which finds a greedy feasible solution to the dual formulation. The algorithm maintains a set of~$C$ of \emph{active terminals}, initially consisting of~$T \setminus \{r\}$. For a terminal~$t \in T$, let~$\mathit{cut}(t)$ be the set of vertices that can reach~$t$ through saturated arcs only. We say that~$t$ induces a \emph{root component} if~$t$ is the only active vertex in~$\mathit{cut}(t)$.

The algorithm~(implicitly) initializes with 0 the variables associated with all Steiner cuts. In each iteration, it picks a vertex~$v \in C$ and checks if~$\mathit{cut}(v)$ is a root component. If it is not, it makes~$v$ inactive and removes it from~$C$; if it is a root component, the algorithm increases~$\pi_{\delta^-(\mathit{cut}(v))}$ until one of its arcs is saturated~(and keeps~$v$ in~$C$). We stop when~$C$ becomes empty, at which point every terminal can be reached from the root using only saturated arcs.

Each iteration takes~$O(|E|)$ time to check a root component. There are~$O(|T|)$ unsuccessful iterations, since each reduces the number of active vertices. A successful iteration saturates at least one arc and increases the size~(number of vertices) of at least one root component. There can be at most $\min\{|V||T|, |E|\}$ such iterations, bounding the total running time by~$O(|E| \min\{|V||T|, |E|\})$~\cite{Dui93}.

\subsection{Our Implementation}
\label{sec:engineering}

We now describe details of our implementation of Wong's algorithm that are crucial to its good performance in practice.

\paragraph{Processing a Root Component.} Once a vertex~$v$ is picked at the beginning of an iteration, we process its~(potential) root component in three passes.

The first pass performs a graph search~(we use BFS) from~$v$ following only saturated incoming arcs. If the search hits another active vertex, the iteration stops immediately:~$v$ does not define a root component. Otherwise, the search finishes with two data structures: a set~$S$ consisting of all vertices in~$\mathit{cut}(v)$, and a list~$L$ containing all unsaturated arcs~$(a,b)$ such that~$a \not\in \mathit{cut}(v)$ and~$b \in \mathit{cut}(v)$. To ensure both structures can be built during the BFS, we allow~$L$ to also contain unsaturated arcs~$(a,b)$ such that both~$a$ and~$b$ belong to~$\mathit{cut}(v)$. These may appear because, when the BFS scans~$b$, it may not know yet whether its neighbor~$a$ will eventually become part of~$\mathit{cut}(v)$; to be safe, we add~$(a,b)$ to~$L$ anyway.

The second pass traverses~$L$ with two aims: (1) remove from~$L$ all arcs~$(a,b)$ that are invalid~(with~$a \in \mathit{cut}(v)$); and (2) pick, among the remaining arcs, the one with the minimum residual capacity~$\Delta$.

The third pass performs an augmentation by reducing the residual capacity of each arc in~$L$ by~$\Delta$. It also builds a set~$X$ with the tails of all saturated arcs, which will be part of the new root component of~$v$~(after augmentation). 

Note that this three-pass approach requires a single graph search~(in the first pass); the other passes are much cheaper, as they merely traverse arrays.

\paragraph{Selection Rules and Lazy Evaluation.} The bound given by the algorithm depends on which active vertex~(root component) it selects in each iteration. Without loss of generality, we assume each iteration picks the active vertex that minimizes some \emph{score} function. Poggi de Arag\~{a}o et al.~\cite{PUW01}~(see also~\cite{Wer01}) found that using the number of incident arcs as the score works well in practice. Polzin and Vahdati~\cite{Pol03,Vah03} show that a related~(but coarser) measure, the number of vertices in the component, also works well. For either score function~(and others), the main challenge is to maintain scores efficiently for all root components during the algorithm, since augmenting on one root component may affect several others.

For efficiency, we focus on \emph{nondecreasing} score functions: as the root component grows, its score can either increase or stay the same. This allows us to use \emph{lazy evaluation}. We maintain each active vertex~$v$ in a priority queue, with a priority~$\sigma(v)$ that is a lower bound on the score of its root component. Each round of the algorithm removes the minimum element~$t$ from the queue. It then verifies~(using the procedure above) if~$t$ defines a root component; if it does not, we just discard~$t$. Otherwise, we perform the corresponding augmentation as long as the actual score is not higher than the priority of the second element in the queue. Finally, we reinsert~$t$ into the priority queue.

When we do augment, recomputing the new exact score of~$t$ can be expensive. Instead, we leverage the fact that we have the set~$X$ of the tails of all arcs saturated during the augmentation. We update the score assuming the vertices in the root component are the union of~$X$ and the original vertices~(at the beginning of the iteration). Although we may miss some vertices, this is relatively cheap to compute and provides a tighter lower bound on the actual score.

Since the number of arcs incident to a root component may decrease, we cannot use it as score function. Instead, we use a refined version of the number of vertices in the root component. Given a component~$c$, let~$\mbox{vc}(c)$ be its number of vertices and let~$\mbox{deg}(c)$ be the sum of their in-degrees. We use~$\mbox{deg}(c) - (\mbox{vc}(c) - 1)$ as the score. This is an upper bound on the number of incoming arcs on the component~(the~$\mbox{vc}(c) - 1$ term discards arcs in a spanning tree of the component, which must exist). This function is nondecreasing, as cheap to compute as the number of vertices, and gives a better estimate on the number of incoming arcs.

\paragraph{Eager evaluation.} Even with lazy evaluation, we may process a root component multiple times before actually performing an augmentation~(or discarding the component). To make the algorithm more efficient, we also use \emph{eager evaluation}: after removing a component from the priority queue, we sometimes perform an augmentation even its real score does not match the priority in the queue. More precisely, as long as the actual score is no more than 25\% higher than the priority, the augmentation is performed. This has almost no effect on solution quality but makes the algorithm significantly faster. Note that any constant factor~(including 25\%) implies a logarithmic bound on the number of times any component can be reevaluated during the algorithm.

\paragraph{Last component.} Typically, the initial cuts found by the dual ascent algorithm have very few arcs, while later ones are much denser. In particular, when there is only one active vertex~$v$ left, we may have to perform several expensive augmentations until it becomes reachable from the root. We can obtain the same bounds faster by dealing with this case differently: we run~(forward) Dijkstra's algorithm from the root to~$v$, using reduced costs as arc lengths. We then use a linear pass to update the reduced costs of the remaining arcs appropriately~\cite{Wer01}.

\subsection{Branch-and-Bound}
\label{sec:bb}

To test the effectiveness of our dual ascent algorithm, we implemented a simple branch-and-bound procedure. We follow the basic principles of most previous work~\cite{Won84,PUW01,Wer01,Pol03,Vah03}, using dual ascent for lower bounds and branching on vertices. The remainder of this section describes other features of our implementation.

The dual ascent root is picked uniformly at random~(among the terminals) and independently for each node of the branch-and-bound tree.

To find primal~(upper) bounds, we run the SPH heuristic on the~(directed) subgraph consisting only of arcs saturated by the dual ascent procedure, using the same root. We then run a single pass of the Steiner vertex insertion and elimination local search procedures~(using all edges, not just saturated ones).

We branch on the vertex that has maximum degree in the primal solution found in the current branch-and-bound node. In case of ties, we look beyond the current primal solution and prefer vertices that maximize the sum of incoming saturated arcs, outgoing saturated arcs, and total degree. Remaining ties are broken at random. If~$v$ is the chosen vertex, we remove it from the graph on the ``zero'' side, and make it a terminal on the ``one'' side. We traverse the branch-and-bound tree in DFS order, visiting the ``one'' side first. This tends to find good primal solutions quicker than other approaches we tried.

We can eliminate an arc~$(u,v)$ if its reduced cost is at least as high as the difference between the best known primal solution and the current dual solution. We actually take the \emph{extended reduced cost}~\cite{PV01b}, which also considers the distance~(using reduced costs) from the root to~$u$. Since the root can change between nodes in the branch-and-bound tree, we only eliminate an edge if both corresponding arcs~(directions) could be fixed by reduced cost. If we fix at least~$|E|/5$ edges in a node, we create a single child node rather than branching.

\section{Improving Robustness}
\label{sec:robust}

While the algorithm we defined in Section~\ref{sec:meta} works well on many graph classes, there are still opportunities to make it more robust~(compared to other approaches) for very easy or very hard instances. Section~\ref{sec:gms} shows how the lower bounds described in Section~\ref{sec:lower} allow our heuristic to stop sooner. Section~\ref{sec:preprocessing} describes some basic preprocessing techniques to reduce the size of the graph on certain classes of instances. Finally, Section~\ref{sec:twophase} describes a two-level version of the multistart algorithm that achieves greater diversification on longer runs.

\subsection{Guarded Multistart}
\label{sec:gms}

Pure heuristics~(such as the multistart algorithm described in Section~\ref{sec:meta}) can be wasteful. Because our heuristic cannot prove that the best solution it found is optimal~(even if it actually is), it cannot stop until it completes its scheduled number of iterations. Ideally, on easy instances, we would like to stop sooner.

To that end, we propose a \emph{Guarded Multistart}~(GMS) algorithm. It runs two threads in parallel: the first runs our standard multistart algorithm, while the second runs the branch-and-bound routine from Section~\ref{sec:bb}. The algorithm terminates as soon as either the multistart thread completes its iterations, or the branch-and-bound thread proves that the incumbent solution is optimal. Communication between threads is limited: the threads inform one another about termination and share the best incumbent solution.

The benefits of this approach are twofold. First, as already mentioned, on easy instances the branch-and-bound algorithm can often prove optimality well before the scheduled number of multistart iterations is reached, making the algorithm faster. Moreover, there are cases in which the branch-and-bound algorithm can find better solutions by itself, leading to better quality as well.

Of course, these advantages are not free: the cycles spent on the branch-and-bound computation could have been used for the multistart itself. On harder instances, we can only afford to perform roughly half as many iterations in the same CPU time. To make this problem less pronounced, we use a simple heuristic to detect cases in which the branch-and-bound computation is obviously unhelpful: if its depth reaches 100, we stop it and proceed only with the multistart computation, saving CPU time.

Another potential drawback is nondeterminism: due to scheduling, different runs of our algorithm may find different results. Although one could make the algorithm deterministic~(by carefully controlling when communication occurs), it is not clear this is worth the extra effort and complexity.

\subsection{Reduction Techniques}
\label{sec:preprocessing}

To be competitive with state-of-the-art algorithms on standard benchmark instances, we must deal with ``easy'' inputs effectively. We thus implemented some basic reduction~(preprocessing) techniques that transform the input into a potentially much smaller instance with the same solution.

In particular, we delete non-terminal vertices of degree one~(alongside their incident edges). Also, if there is a non-terminal vertex~$v$ with exactly two neighbors,~$u$ and~$w$, we replace edges~$(u,v)$ and~$(v,w)$ with a single edge~$(u,w)$ with cost~$\cost(u,w) = \cost(u,v) + \cost(v,w)$. Finally, we implemented a limited version of the \emph{Bottleneck Steiner Distance}~\cite{DV89} test, which states that an edge~$(u,v)$ can be removed from the graph if there is a~(\emph{bottleneck}) path~$P_{uv}$ between~$u$ and~$v$ such that (1)~$P_{uv}$ excludes~$(u,v)$ and (2) every subpath of~$P_{uv}$ without an internal terminal has length at most~$\cost(u,v)$. Identifying all removable edges can be expensive, so we restrict ourselves to common~(and cheap) special cases: paths with up to two vertices~(if both~$u$ and~$v$ have small degree) as well as some longer paths within a spanning tree of the graph. Both can be seen as restricted versions of existing algorithms~\cite{Pol03,Vah03}. See Appendix~\ref{sec:bottleneck} for details.

\subsection{Two-Phase Multistart}
\label{sec:twophase}

Even with all the measures we take to increase diversification, the algorithm can still be strongly influenced by the first few solutions it finds, since they will be heavily used during cascaded combination. If the algorithm is unlucky in the choice of the first few solutions, it may be unable to escape a low-quality local minimum.

When the number~$M$ of iterations is large~(in the thousands), we obtain more consistent results with a two-phase version of our algorithm. The first phase independently runs the standard algorithm four times, with~$M/8$ iterations each. The second phase runs the standard algorithm with~$M/2$ iterations, but starting from a pool of elite solutions obtained from the union of the four pools created in the first phase. Note that the combined size of all pools in the first phase is~$4 \sqrt{M/16} = \sqrt{M}$, while the second-phase pool can hold only~$\sqrt{M/4} = \sqrt{M}/2$ solutions. We thus take the elite solutions from the first phase in random order and try to add them to the~(initially empty) final pool, using the criteria outlined in Section~\ref{sec:pool} to decide which solutions are kept.

We call this two-phase version of our multistart algorithm MS2. Since it has multiple independent starts, it is less likely than MS to be adversely influenced by a particularly bad initial solution.

\section{Experiments}
\label{sec:experiments}

\begin{table}[t!]
\centering
\caption{Classes of instances tested.\label{tab:families}}{
\begin{tabular}{lll}
\toprule
\textsc{class} & \phantom{a}\textsc{series} & \phantom{a}\textsc{description} \\
\midrule
\class{euclidean} & \phantom{a}\sseries{x} \sseries{p4e} \sseries{p6e} & \phantom{a}Euclidean costs \cite{CGR92,KM98}\\
\class{fst} & \phantom{a}\sseries{es*fst} \series{tspfst} \series{cph14} & \phantom{a}reduced geometric, L1 costs~\cite{WWZ00,dimacs11,HY13}\\
\class{hard} & \phantom{a}\sseries{bip} \sseries{cc} \sseries{hc} \sseries{sp} & \phantom{a}synthetic hard instances \cite{RPRUW01,KMV00}\\
\class{incidence} & \phantom{a}\sseries{i080} \sseries{i160} \sseries{i320} \sseries{i640} & \phantom{a}random graphs, incidence costs \cite{Dui93}\\
\class{r} & \phantom{a}\sseries{1r} \sseries{2r} & \phantom{a} 2D and 3D cross-grid graphs \cite{Fre97c}\\
\class{random} & \phantom{a}\sseries{b} \sseries{c} \sseries{d} \sseries{e} \sseries{mc} \sseries{p4z} \sseries{p6z} & \phantom{a}graphs with random costs \cite{Bea90,CGR92,KM98}\\
\class{vienna} & \phantom{a}\sseries{gori} \sseries{gadv} \sseries{isim} \sseries{iadv} & \phantom{a}road networks \cite{LLLPR14}\\
\class{vlsi} & \phantom{a}\sseries{alue} \sseries{alut} \sseries{dmxa} \sseries{diw} \sseries{gap} \sseries{lin} \sseries{msm} \sseries{taq} & \phantom{a}planar grid graphs with holes \cite{KM98,KMV00}\\
\class{wrp} & \phantom{a}\sseries{wrp3} \sseries{wrp4} & \phantom{a}group Steiner grid instances \cite{ZR00}\\
\bottomrule
\end{tabular}}
\end{table}

We implemented all algorithms in C++ and compiled them using Visual Studio 2013~(optimizing for speed). We ran our experiments on a machine two 3.33\,GHz Intel Xeon X5680 processors running Windows 2008R2 Server with 96\,GB of DDR3-1333 RAM. This machine scores 388.914 according to the benchmark code made available for the 11th DIMACS Implementation Challenge~(\url{http://dimacs11.cs.princeton.edu/downloads.html}). All runs we report are sequential, except those of the Guarded Multistart algorithm, which use two cores. In every case, we report total CPU times, i.e., the sum of the times spent by each CPU involved in the computation.

We evaluate all instances available by August 1, 2014 from the 11th DIMACS Implementation Challenge~\cite{dimacs11}. We group each original series into \emph{classes}, as shown in Table~\ref{tab:families}~(augmented from~\cite{UW12}). More detailed information about the dimensions of the instances in each series can be found in Table~\ref{tab:sizes}, in the appendix. Most instances are available from the SteinLib~\cite{KMV00}, with two exceptions: \instance{cph14}~(graphs obtained from rectilinear problems~\cite{HY13}) and \instance{vienna}~(road networks from telecommunication applications~\cite{LLLPR14}). When evaluating solution quality, we compare to the best results published by August 1, 2014~(listed at \url{http://dimacs11.cs.princeton.edu/downloads.html}).

\subsection{Multistart}

In our first experiment, we ran the default version of our multistart algorithm on all instances from the DIMACS Challenge~\cite{dimacs11}. Recall that this version is not guarded~(no branch-and-bound), but uses the lightweight preprocessing routine. We vary the number of multistart iterations from 1 to 256~(by factors of 4). Table~\ref{tab:msclass} shows average running times~(in seconds) and percent errors relative to the best known solutions. (Results aggregated by series can be found in Tables~\ref{tab:msserieserror} and~\ref{tab:msseriestime}, in the appendix.) Because percent errors are very small for the \instance{wrp} instances~(a side effect of a reduction from the Group Steiner Tree problem), to improve readability we show the error for \instance{wrp} multiplied by~$1000$ in this and other tables. The special entries are marked in \textsl{slanted font}.

\begin{table}[t!]
\centering
\caption{Multistart Algorithm: Average running time in seconds and average percent error relative to the best known solutions, with number of iterations varying from 1 to 256. (Errors are multiplied by~$10^3$ for \instance{wrp}; for example, entry 0.496 actually means~$0.000496\%$.)}
\label{tab:msclass}
\begin{tabular}{l rrrrr rrrrr}
\toprule
\multicolumn{1}{c}{} & \multicolumn{5}{c}{\tabhead{time} [s]} & \multicolumn{5}{c}{\tabhead error [\%]}\\
\cmidrule(lr){2-6} \cmidrule(l){7-11}
\tabhead{class} & 1 & 4 & 16 & 64 & 256 & 1 & 4 & 16 & 64 & 256 \\
\midrule
\instance{euclidean} & 0.005 & 0.011 & 0.036 & 0.135 & 0.520 & 0.155 & 0.009 & 0.000 & \optsol & \optsol \\
\instance{fst} & 0.014 & 0.089 & 0.433 & 1.862 & 7.560 & 0.540 & 0.198 & 0.069 & 0.027 & 0.012 \\
\instance{hard} & 0.047 & 0.340 & 1.855 & 7.626 & 30.377 & 4.499 & 2.726 & 1.607 & 0.918 & 0.492 \\
\instance{incidence} & 0.146 & 0.445 & 1.757 & 7.253 & 29.496 & 2.608 & 0.967 & 0.374 & 0.141 & 0.043 \\
\instance{r} & 0.012 & 0.042 & 0.166 & 0.688 & 2.698 & 4.292 & 1.628 & 0.445 & 0.133 & 0.056 \\
\instance{random} & 0.009 & 0.027 & 0.104 & 0.424 & 1.687 & 1.104 & 0.217 & 0.049 & 0.014 & 0.009 \\
\instance{vienna} & 0.532 & 3.880 & 19.822 & 85.522 & 356.237 & 0.074 & 0.035 & 0.017 & 0.008 & 0.000 \\
\instance{vlsi} & 0.045 & 0.234 & 1.086 & 4.654 & 18.902 & 0.809 & 0.271 & 0.082 & 0.028 & 0.011 \\
\instance{wrp} & 0.015 & 0.080 & 0.353 & 1.480 & 5.932 & \textsl{0.496} & \textsl{0.129} & \textsl{0.026} & \textsl{0.004} & \textsl{0.001} \\

\bottomrule
\end{tabular}
\end{table}

Our algorithm is quite effective. With as little as 16 multistart iterations, the average error rate is below 0.5\% on all classes except \instance{hard}, which consists of adversarial synthetic instances. With 256 iterations, the average error falls below 0.5\% for \instance{hard}, and below 0.06\% for all other classes. Average running times are still quite reasonable: the only outlier is \instance{vienna}, which has much bigger graphs on average~(see Table~\ref{tab:sizes}, in the appendix). These instances are relatively easy: a single iteration is enough to find solutions that are on average within 0.1\% of the best known.

\paragraph{Guarded Multistart.} Recall that our \emph{Guarded Multistart}~(GMS) algorithm runs two parallel threads, the first with the standard multistart and the second with a branch-and-bound algorithm. The algorithm stops as soon as either the multistart algorithm completes its scheduled number of iterations or when the branch-and-bound procedure proves the solution is optimal. We tested GMS with 2, 8, 32, and 128 multistart iterations. Table~\ref{tab:gmsclass} reports the error rates and average running times~(see also Tables~\ref{tab:gmsserieserror} and \ref{tab:gmsseriestime}, in the appendix). For consistency, we report total CPU times; since GMS uses two cores, the actual wall-clock time is lower.

\begin{table}[b!]
\centering
\caption{Guarded Multistart: Average CPU time in seconds and average percent error relative to the best known solutions. (Errors are multiplied by~$10^3$ for \instance{wrp}.)}
\label{tab:gmsclass}
\begin{tabular}{l rrrr rrrr}
\toprule
\multicolumn{1}{c}{} & \multicolumn{4}{c}{\tabhead{time} [s]} & \multicolumn{4}{c}{\tabhead{error} [\%]}\\
\cmidrule(lr){2-5} \cmidrule(l){6-9}
\tabhead{class} & 2 & 8 & 32 & 128 & 2 & 8 & 32 & 128 \\
\midrule
\instance{euclidean} & 0.009 & 0.010 & 0.011 & 0.010 & 0.000 & \optsol & \optsol & \optsol \\
\instance{fst} & 0.063 & 0.398 & 1.757 & 7.235 & 0.282 & 0.095 & 0.040 & 0.018 \\
\instance{hard} & 0.304 & 1.704 & 7.313 & 30.016 & 3.094 & 1.709 & 1.070 & 0.648 \\
\instance{incidence} & 0.403 & 1.012 & 2.617 & 4.400 & 0.360 & 0.065 & 0.030 & 0.019 \\
\instance{r} & 0.036 & 0.117 & 0.399 & 1.456 & 0.876 & 0.346 & 0.118 & 0.056 \\
\instance{random} & 0.021 & 0.050 & 0.127 & 0.481 & 0.119 & 0.044 & 0.014 & 0.004 \\
\instance{vienna} & 3.663 & 19.604 & 87.347 & 358.916 & 0.058 & 0.024 & 0.011 & 0.003 \\
\instance{vlsi} & 0.279 & 1.144 & 4.497 & 18.427 & 0.364 & 0.067 & 0.035 & 0.012 \\
\instance{wrp} & 0.068 & 0.331 & 1.424 & 5.700 & \textsl{0.175} & \textsl{0.033} & \textsl{0.008} & \textsl{0.002} \\

\bottomrule
\end{tabular}
\end{table}

The CPU time spent by GMS with~$i$ iterations cannot be~(by design) much worse that the unguarded algorithm~(MS) with~$2i$ iterations. A comparison of Tables~\ref{tab:msclass} and~\ref{tab:gmsclass} shows that running times are indeed similar for several classes, such as \instance{hard}, \instance{vlsi}, and \instance{wrp}. But GMS can stop much sooner on ``easy'' instances, when its branch-and-bound portion can quickly prove the optimality of the incumbent. For \instance{random}, \instance{incidence}, and especially \instance{euclidean}, GMS is faster than MS.

The relative solution quality of the two variants also depends on the type of instance. For classes such as \instance{hard} and \instance{wrp}, the guarded variant finds slightly worse results~(for the same amount of CPU time), since most of the useful computation is done by the multistart portion of the algorithm, which has fewer iterations to work with. For a few classes~(such as \instance{incidence}), the guarded version actually finds much better solutions, thanks to the branch-and-bound portion of the algorithm. In most cases, the difference is quite small. On balance, the guarded version is more robust and should be used unless there is reason to believe the branch-and-bound portion will be ineffective.

\paragraph{Comparison.}

Table~\ref{tab:gmspv} compares Guarded Multistart~(GMS) against the three state-of-the-art heuristics presented by Polzin and Vahdati~\cite{Pol03,Vah03}: \textsc{prune}, \textsc{ascent\&prune}, and \textsc{slack-prune}. In particular, as reported by Polzin and Vahdati~\cite{Pol03,Vah03}, they dominate the multistart approach by Ribeiro et al.~\cite{RUW02}.

Since the three algorithms have very different time/quality tradeoffs, we report results for GMS with 1, 8, 32, and 128 iterations. For consistency with how the results are reported in~\cite{Pol03,Vah03}, Table~\ref{tab:gmspv} shows the \instance{lin} series separately from the remaining \instance{vlsi} instances. Running times for their algorithms are scaled~(divided by 6.12) to match our machine; Appendix~\ref{sec:scaling} explains this factor.

\begin{table}[t!]
\centering
\caption{Comparison of various heuristics. For \instance{wrp} instances, all errors are multiplied by~$10^3$.}
\scriptsize
\label{tab:gmspv}
\begin{tabular}{l rr rr r@{\hspace{2.5pt}}r rr rr r@{\hspace{2.5pt}}r r@{\hspace{2.5pt}}r}
\toprule
\multicolumn{1}{c}{} & \multicolumn{6}{c}{\tabhead{previous algorithms}~\cite{Pol03,Vah03}} & \multicolumn{8}{c}{\tabhead{guarded multistart}} \\
\cmidrule(lr){2-7} \cmidrule(l){8-15}
\multicolumn{1}{c}{} & \multicolumn{2}{c}{\tabhead{prune}} & \multicolumn{2}{c}{\tabhead{ ascend\&prune}} & \multicolumn{2}{c}{\tabhead{slack-prune}} & \multicolumn{2}{c}{\tabhead GMS1} & \multicolumn{2}{c}{\tabhead GMS8} & \multicolumn{2}{c}{\tabhead GMS32} & \multicolumn{2}{c}{\tabhead GMS128} \\
\cmidrule(lr){2-3} \cmidrule(lr){4-5} \cmidrule(lr){6-7} \cmidrule(lr){8-9} \cmidrule(lr){10-11} \cmidrule(lr){12-13} \cmidrule(l){14-15}
\tabhead{group} & \tabhead{err.} & \tabhead{time} & \tabhead{err.} & \tabhead{time} & \tabhead{err.} & \tabhead{time} & \tabhead{err.} & \tabhead{time} & \tabhead{err.} & \tabhead{time} & \tabhead{err.} & \tabhead{time} & \tabhead{err.} & \tabhead{time} \\
\midrule
\instance{1r} & 1.360 & 0.021 & 1.030 & 0.011 & \optsol & 0.036 & 0.662 & 0.014 & 0.080 & 0.056 & \optsol & 0.147 & \optsol & 0.407 \\
\instance{2r} & 1.420 & 0.044 & 1.590 & 0.026 & \optsol & 1.783 & 2.363 & 0.030 & 0.612 & 0.177 & 0.236 & 0.651 & 0.113 & 2.505 \\
\instance{d} & 0.070 & 0.016 & 0.020 & 0.011 & \optsol & 0.023 & 0.271 & 0.018 & 0.077 & 0.050 & \optsol & 0.073 & \optsol & 0.092 \\
\instance{e} & 0.310 & 0.064 & 0.130 & 0.041 & \optsol & 0.268 & 0.413 & 0.048 & 0.107 & 0.192 & 0.076 & 0.577 & 0.024 & 2.483 \\
\instance{es10000fst} & 1.110 & 1.235 & 0.670 & 5.046 & 0.380 & 343.446 & 0.651 & 1.266 & 0.374 & 28.187 & 0.218 & 123.472 & 0.151 & 519.184 \\
\instance{es1000fst} & 1.010 & 0.093 & 0.530 & 0.062 & 0.190 & 3.034 & 0.654 & 0.085 & 0.229 & 1.165 & 0.118 & 4.980 & 0.045 & 20.540 \\
\instance{i080} & 1.150 & 0.011 & 1.650 & 0.003 & 0.060 & 0.070 & 0.349 & 0.008 & \optsol & 0.011 & \optsol & 0.011 & \optsol & 0.011 \\
\instance{i160} & 1.970 & 0.051 & 1.690 & 0.011 & 0.100 & 0.275 & 0.873 & 0.028 & 0.030 & 0.058 & 0.003 & 0.085 & \optsol & 0.088 \\
\instance{i320} & 2.840 & 0.266 & 1.810 & 0.049 & 0.140 & 1.219 & 1.099 & 0.140 & 0.075 & 0.362 & 0.040 & 0.658 & 0.027 & 1.566 \\
\instance{lin} & 1.440 & 0.247 & 0.760 & 0.178 & 0.040 & 25.162 & 0.876 & 0.510 & 0.114 & 2.590 & 0.052 & 10.327 & 0.021 & 43.031 \\
\instance{mc} & 1.700 & 0.008 & 1.010 & 0.007 & 0.420 & 0.155 & 0.192 & 0.009 & 0.181 & 0.031 & \optsol & 0.072 & \optsol & 0.085 \\
\instance{tspfst} & 0.420 & 0.034 & 0.310 & 0.062 & 0.040 & 5.230 & 0.350 & 0.040 & 0.096 & 0.469 & 0.032 & 2.158 & 0.016 & 8.984 \\
\instance{vlsi} & 0.390 & 0.054 & 0.350 & 0.065 & 0.004 & 1.172 & 0.555 & 0.064 & 0.053 & 0.683 & 0.030 & 2.637 & 0.009 & 10.579 \\
\instance{wrp3} & \textsl{0.6} & 0.025 & \textsl{0.3} & 0.023 & \textsl{0.03} & 2.907 & \textsl{0.218} & 0.039 & \textsl{0.039} & 0.403 & \textsl{0.012} & 1.758 & \textsl{0.003} & 7.074 \\
\instance{wrp4} & \textsl{70} & 0.016 & \textsl{0.6} & 0.016 & \textsl{0.06} & 1.011 & \textsl{0.531} & 0.027 & \textsl{0.027} & 0.259 & \textsl{0.003} & 1.085 & \optsol & 4.303 \\
\instance{x} & 0.170 & 0.072 & \optsol & 0.047 & \optsol & 0.038 & 0.006 & 0.024 & \optsol & 0.045 & \optsol & 0.045 & \optsol & 0.045 \\

\bottomrule
\end{tabular}
\end{table}

The table shows that the algorithms have very different profiles. Both \tabhead{prune} and \tabhead{ascend\&prune} are quite fast, with running times comparable to GMS1, which runs a single multistart iteration. They provide much better solutions on series \instance{d} and \instance{e}, whereas GMS1 is significantly better on \instance{1r}, \instance{i080}, \instance{i160}, \instance{i320}, and \instance{x}. Error rates are usually within a factor of two of one another otherwise.

The \textsc{slack-prune} algorithm usually finds better solutions, but takes much longer; it should then be compared with GMS with a few dozen iterations. The \textsc{slack-prune} approach is superior when advanced reduction techniques~(exploiting small duality gaps) work very well: \instance{2r}, \instance{e}, \instance{vlsi}, are good examples. When these techniques are less effective, our algorithm dominates: see \instance{es10000fst}, \instance{i080}, \instance{i160}, \instance{i320}, \instance{mc}, and \instance{wrp}, for example. Performance is comparable for several cases in between, such as \instance{es1000fst}, \instance{lin}, or \instance{tspfst}.

Unfortunately, Polzin and Vahdati~\cite{Pol03,Vah03} only report results for series in which all optimal solutions are known, which consist mostly of small inputs or instances for which reduction techniques work well. The fact that GMS is competitive even in the absence of very hard instances is encouraging. It shows that, although reduction and dual-based techniques are powerful, primal heuristics based on local search~(the core of our approach) are essential for a truly robust algorithm.

\begin{wraptable}[22]{r}{.41\textwidth}
\centering
\caption{Results for three MS variants on 41 open SteinLib instances: (geometric) mean time in seconds and average percent error relative to the best known solution.}
\label{tab:longagg}
\begin{tabular}{l@{\hspace{-.3em}}r@{\hspace{.2em}}r@{\hspace{.2em}}r}
\toprule
\tabhead{method} & \tabhead{iter.} & \tabhead{time [s]} & \tabhead{err.\,[\%]} \\
\midrule
MS & 1024 & $128.0$ & $0.407$ \\
& 4096 & $520.7$ & $0.334$ \\
& 16384 & $2094.8$ & $0.213$ \\
& 65536 & $8474.3$ & $0.096$ \\
& 262144 & $35056.4$ & $0.107$\vspace{1 pt} \\
MS2 & 1024 & $126.8$ & $0.357$ \\
& 4096 & $507.6$ & $0.150$ \\
& 16384 & $2022.3$ & $0.050$ \\
& 65536 & $8221.0$ & $-0.020$ \\
& 262144 & $32609.4$ & $-0.094$\vspace{1 pt} \\
MSK & 64 & $271.6$ & $0.623$ \\
& 256 & $1096.3$ & $0.360$ \\
& 1024 & $4359.1$ & $0.218$ \\

\bottomrule
\end{tabular}
\end{wraptable}

\paragraph{Long Runs.} To test the scalability of our algorithm, we consider the 41 SteinLib instances with no published proof of optimality by August 1, 2014. We consider three versions of our algorithm. The baseline is MS, the multistart algorithm described in Section~\ref{sec:meta}. MS2 is the two-phase version of MS, as described in Section~\ref{sec:twophase}. Finally, MSK augments plain MS by also using the \emph{key-vertex insertion} local search implemented as calls to the SPH algorithm~(as proposed in~\cite{PRUW01}); it is very expensive, but can find better results. None of these variants is guarded, since branch-and-bound is ineffective on hard instances. To find near-optimal solutions, we test up to 262144~($2^{18}$) iterations~(1024 for MSK).

For each variant~(and number of iterations), Table~\ref{tab:longagg} shows results aggregated over all 41 instances: (geometric) mean time in seconds and average error with respect to the best published solutions.

With 1024 iterations, MSK finds better results than other variants, but is much slower: increasing the number of iterations of either MS or MS2 to 16384 is cheaper and leads to better solutions. Unsurprisingly, MS and MS2 have comparable running times for the same number of iterations. As argued in Section~\ref{sec:twophase}, increasing the number of iterations is more effective for MS2 than for MS. In fact, the average solution quality for MS does not even improve when we increase the number of iterations from~$65536$ to~$262144$. By starting from four independent sets of solutions, MS2 is less likely to be confined to a particularly bad region of the search space.

Considering all 13 runs from Table~\ref{tab:longagg}~(five runs each for MS and MS2, and three for MSK), there were only six cases~(out of 41) for which we could not at least match the best bound published by August 1, 2014. (See Table~\ref{tab:longsol}, in the appendix.) In 21 cases, we found a strictly better solution. Most of these were found by MS2~(see Table~\ref{tab:ms2sol}, in the appendix); MSK was better only for \instance{cc11-2u} and \instance{cc12-2u}.

\subsection{Branch-and-Bound}
\label{sec:expbb}

For completeness, we now consider the effectiveness of our branch-and-bound procedure as a standalone exact algorithm. Unlike our heuristics, it is not robust. There are some graphs~(such as large \instance{vlsi} or \instance{fst} instances) for which our algorithm will not produce a good solution in reasonable time, let alone prove its optimality. On large instances with small duality gaps, algorithms based on dual ascent are generally not competitive with exact algorithms that use linear programming.

We thus focus on small instances with large duality gaps. Series \instance{i080}, \instance{i160}, and \instance{i320} have been solved to optimality~\cite{KM98,PUW01}, as have 95 of 100 instances from \instance{i640}~\cite{Pol03,Vah03}. We also consider all solved instances from the \textsf{bip} and \textsf{cc} series. As shown in Tables~\ref{tab:i080}, \ref{tab:i160}, \ref{tab:i320}, \ref{tab:i640}, and \ref{tab:puc}~(in the appendix), our method could solve every such instance in less than two hours; most took fractions of a second.

\begin{table}[t!]
\centering
\caption{Performance of our branch-and-bound algorithm on select hard instances, in comparison with the exact algorithm by Polzin and Vahdati Daneshmand~\cite{Pol03,Vah03}. }
\label{tab:bbpv}
\begin{tabular}{lrrrrrrr}
\toprule
\multicolumn{2}{c}{} & \multicolumn{3}{c}{\tabhead{average time}} & \multicolumn{3}{c}{\tabhead{mean time}} \\
\cmidrule(lr){3-5} \cmidrule(l){6-8}
\tabhead{series} & \tabhead{count} & \tabhead{ours} & \cite{Pol03,Vah03} & \tabhead{ratio} & \tabhead{ours} & \cite{Pol03,Vah03} & \tabhead{ratio} \\
\midrule
\instance{i160} & 100 & 0.042 & 0.384 & 9.2 & 0.009 & 0.066 & 7.4 \\
\instance{i320} & 100 & 13.195 & 211.834 & 16.1 & 0.069 & 0.428 & 6.2 \\
\instance{i640} & 95 & 176.852 & 1363.153 & 7.7 & 0.468 & 2.492 & 5.3 \\
\instance{cc} & 8 & 214.706 & 5384.671 & 25.1 & 1.102 & 32.404 & 29.4 \\
\instance{bip} & 2 & 225.316 & 572.083 & 2.5 & 224.113 & 571.383 & 2.5 \\

\bottomrule
\end{tabular}
\end{table}

For perspective, Table~\ref{tab:bbpv} compares our exact algorithm against the state-of-the-art approach for such instances, due to Polzin and Vahdati~\cite{Pol03,Vah03}. The table has all instances they solved from series \instance{i160}, \instance{i320}, \instance{i640}, \instance{cc}, and \instance{bip}. For each series, we show the number of instances tested, our average time in seconds, the average time of their method~(divided by 6.12 to match our machine), and the ratio between them. The remaining columns use geometric means instead of averages.

Our method is quite competitive for these instances. Running times are comparable for \instance{bip} instances and we are faster for other graph classes. The relative difference is higher when we consider averages rather than mean times, indicating that our advantage is greater on harder instances~(which have a more pronounced effect on the average). This confirms that the engineering effort outlined in Section~\ref{sec:lower} does pay off.

We stress, however, that Table~\ref{tab:bbpv} contains only a very small~(and not particularly representative) subset of all instances tested. Because Polzin and Vahdati use linear programming and advanced reduction techniques, there are several classes of instances~(such as \instance{vlsi}) that they can easily solve but we cannot. This is true for other algorithms as well~\cite{HSV14}. Even for some of the instances in Table~\ref{tab:bbpv}, the algorithm by Polzin and Vahdati has become more competitive since its initial publication. They recently reran~\cite{PV14} their original algorithm on a newer machine with different sets of parameters and using an up-to-date version of CPLEX. For series \instance{i160}, \instance{i320}, \instance{i640}, and \instance{bip}, their~(scaled) average running times in seconds are now 0.08, 65.2, 169.3, and 389.4. These improvements~(of at least a factor of three) bring their algorithm closer to ours. For \instance{cc}, however, scaled average times are only slightly better~(4890 instead of 5385), which makes our method still more than 20 times faster.

Finally, we note that our branch-and-bound algorithm could prove that 35535 is the optimal solution for \instance{i640-313}, a formerly open \instance{incidence} instance. On a machine about 28\% faster than the one we used for all other experiments, it took 15.16 days and visited 7.31 billion branch-and-bound nodes. For this particular run, we gave the algorithm 35536~(just above the optimum) as the initial upper bound and used strong branching~(as described in Section~\ref{sec:strong}, in the appendix).

\section{Conclusion}
\label{sec:conclusion}

We presented a new heuristic approach for the Steiner problem in graphs, based on fast local searches, multistart with diversification, and fast combinatorial algorithms for finding lower bounds. Although the algorithm could be further improved, notably by incorporating more elaborate preprocessing techniques, it is already quite robust. For short runs, it is competitive with any previous approach on a wide variety of instance classes. Moreover, it is scalable: when given more time, it improved the best published solutions~(as of August 1, 2014) for 21 number hard instances from the literature. We note that some bounds have also been improved by other submissions to the 11th DIMACS Challenge~\cite{dimacs11}. According to a summary published\footnote{\url{http://dimacs11.cs.princeton.edu/instances/bounds20140912.txt}} on September 12, 15 of the 21 improved solutions we found remain the best known~(and 13 of those have not been matched by other approaches). Overall, our results show that primal heuristics can be an important component of robust solvers for the Steiner problem in graphs.

\bibliographystyle{shortAbbrv}

\clearpage

\appendix

\section{Appendix}

\subsection{Contest Entry}

The work presented in this article served as the basis in the contest associated with the 11th DIMACS Implementation Challenge~(in Collaboration with ICERM). In the contest, different programs are analyzed according to the quality of all incumbent solutions within a given time limit~$\tau$. The standard version of our multistart algorithm~(as described in the main text) is parameterized by the number of iterations, rather than a time limit. Moreover, the size of the elite pool depends on the number of iterations, which must be known in advance. This section describes some small adaptations we made to our algorithm to account for the time limit~$\tau$ in a meaningful way.

Our entry in the contest is based on the unguarded algorithm described in Section~\ref{sec:meta}, modified to work in three phases.

The first phase runs a subset of the preprocessing routines described in Section~\ref{sec:preprocessing}, restricted to operations that only remove edges. The fact that no shortcuts are added avoids the need to implement edge-mapping routines.

The second phase runs the standard multistart iteration with a \emph{single iteration}, which consists of the constructive algorithm~(SPH) followed by local search. Unlike our standard algorithm, here we use actual edge weights, without perturbation. Both the constructive solution and the local optimum~(if better) are reported as incumbent solutions.

The third phase is a standard run of the multistart algorithm described in Section~\ref{sec:meta}~(starting from an empty elite pool). It sets the size of the elite pool to~$\sqrt{\hat{M}/2}$, where~$\hat{M}$ is an estimate on the number of iterations the algorithm can run within the time budget~$\tau$. Let~$\tau_1$ be the running time of the second phase~(above). We set~$\hat{M} = \tau / (2.5 \tau_1)$, meaning that we estimate that a standard iteration of the algorithm will take about 2.5 times the first one~(which does not combinations). This constant~(2.5) was found empirically, and the algorithm is not too sensitive to it. We set the maximum number of iterations of the algorithm to~$M = 65536$, but stop sooner if the time limit~$\tau$ is reached. Only incumbent solutions found within the time limit are reported.

\subsection{Scaling}
\label{sec:scaling}

Tables~\ref{tab:gmspv} and \ref{tab:bbpv} compare the running times of our algorithm to those found by Polzin and Vahdati Daneshmand~\cite{Pol03,Vah03}. As previously mentioned, we divide their published times by 6.12 for a fair comparison. This section explains how we arrived at this factor.

We use the scores available at \url{http://www.cpubenchmark.net/singleThread.html} in the absence of more precise benchmarks. The Intel Xeon X5680 processors in our test machine have a score of 1484. The relevant experiments in~\cite{Pol03} use a Sunfire 15000 with 900 MHz SPARC III+ CPUs. Although this machine is not listed in the page above, the authors also executed one of their experiments~(the exact algorithm) on an AMD Athlon XP 1800+ running at 1.53\,GHz, whose score is 477. To compute the relative speed of these two machines, we look at \instance{es10000} and \instance{fnl4461}, the two hardest instances for which results on both machines are given in~\cite{Pol03}. On the Sunfire, Table A.6~\cite{Pol03} states that their code takes 1398.9\,s and 12967.0\,s, respectively. On the AMD, the corresponding times~(given in Table 5.3~\cite{Pol03}) are 758 and 6148. The mean ratio between these values is 1.967, which would give the Sunfire a score of~$477/1.967 \approx 242.5$; our machine is thus~$1484 / 242.5 \approx 6.12$ times faster according to this metric.

For reference, to prove the optimality of \textsf{i640-313} we used an Intel Xeon E5-1620 running at 3.6\,GHz. Its score at \url{http://www.cpubenchmark.net/singleThread.html} is 1911, which is about 28\% higher than our main machine.

To compare our results with those in a recent manuscript by Vahdati and Daneshmand~\cite{PV14}, we divide their running times by 1.269, the ratio between the 11th DIMACS Challenge Benchmark scores of our machine~(388.913664) and theirs~(306.508988).

\subsection{Notes on Local Search}

Uchoa and Werneck~\cite{UW12} state in passing that key-vertex removal ``dominates'' Steiner vertex removal, but this is not strictly true. One can construct local optima for key-vertex removal that can be improved by Steiner vertex removal, since the latter can make an original degree-two vertex (on a key-path) become a key vertex. In practice, however, we observed that such cases are extremely rare.

The original implementations evaluated by Uchoa and Werneck~\cite{UW12} use the C\# programming language; we converted them to C++ for this work. Since C\# is interpreted rather than compiled, our new implementation is often faster by a factor of 3 or more.
Finally, we note that the running times reported in the appendix of \cite{UW12} are for a single iteration (and not to reach the local optimum, as the article mistakenly states).

\subsection{Strong Branching}
\label{sec:strong}

For very long runs of our branch-and-bound algorithm, we found it worth it to use strong branching at higher levels of the search tree. We observed that the variance can be quite large, given the random choices of the dual ascent algorithm. To get a reliable measure of the effect of branching on a vertex~$v$, we would have to perform several simulations.

Instead, we evaluate \emph{sets} of vertices at a time, and infer from that the effect of individual vertices. The idea is simple. First, we pick~$c$ random sets with~$s$ nonterminals each ($c$ and~$s$ are parameters of the algorithm). For each such set, we fix all vertices simultaneously to the same side (0 or 1), run dual ascent on the resulting graph, and note the corresponding dual bound. The final score of each vertex is the average dual bound over all sets that contain~$v$. Higher scores are better.

In practice, we found that the dual bound tends to increase less when eliminating a vertex (side 0) than when making it a terminal (side 1). With that in mind, we consider both cases separately. For the 0 side, we set~$c = 1000$ and~$s = 32 - d$, where~$d$ is the depth in the branch-and-bound tree; for the 1 side, we set~$c = 2000$ and~$s = 10$. Each vertex receives a separate scores~$\sigma_0$ and~$\sigma_1$ for each side (0 and 1), and use~$\sigma = (\sigma_0^3 \cdot \sigma_1)^{1/4}$ as the final score.

We call this approach \emph{scatter branching}. Since it is very expensive, we only use it at small depths in the tree (up to 14) and for very large instances (with billions of branch-and-bound nodes). For \instance{i640-313} (the only instance in this paper for which we used scatter branching), we saw speedups of about a factor of two. Further tuning may lead to better results.

\subsection{Reduction}
\label{sec:bottleneck}

As anticipated in Section~\ref{sec:preprocessing}, we now describe our restricted implementation of the \emph{Bottleneck Steiner Distance} test. Given an edge~$(u,v)$, our goal is to find a bottleneck path between~$u$ and~$v$ that does not use~$(u,v)$ itself. We use two quick heuristics to look for a subset of such paths; while this is correct (we preserve the optimal solution), we miss some opportunities for reduction, leading to a instance that may not be as small as possible.

The first heuristic is simple. If the combined degree of~$u$ and~$v$ is small (up to 20 in our implementation), we scan both vertices looking for a common neighbor~$x$ such that~$\cost(u,x) + \cost(x,v) \leq \cost(u,v)$ (we also check if a parallel~$(u,v)$ edge exists). This heuristic helps with some grid-like graphs, such as VLSI instances.

The second test we implemented is more global and may find more complicated paths. We first use a modified version of Dijkstra's algorithm~\cite{Dij59} to build the Voronoi diagram associated with the terminals of~$G$~\cite{Meh88,PW02,UW12}. For each vertex~$v$, this structure defines~$\vorbase(v)$ (the closest terminal to~$v$),~$\vordist(v)$ (the distance to that terminal), and~$\vorparent(v)$ (the parent edge on the path to~$\vorbase(v)$). By looking at boundary edges of this Voronoi diagram, we compute the MST of the distance network of~$G$~\cite{Meh88,PW02,UW12}.

Let~$E_t$ be the set of edges of this MST, each corresponding to a path in the original graph. Let~$E_f \subseteq E$ be the set of \emph{free edges}, i.e., original edges that are neither parents in the Voronoi diagram nor belong to a path in~$E_t$. We try to eliminate edges in~$E_f$, using~$E_t$ and the Voronoi diagram to find bottleneck paths.

To do so efficiently, we first sort the union of~$E_t$ and~$E_f$ in increasing order of cost, breaking ties in favor of entries in~$E_t$. Then, we initialize a union-find data structure~\cite{Tar83} with~$|V|$ disjoint sets. We then traverse this list in order. Consider edge~$e_i = (u,v)$. If~$e_i \in E_t$, we join groups~$u$ and~$v$ in the union-find data structure. Otherwise (if~$e_i \in E_f$) we delete~$e_i$ if all of the following three conditions hold: (1)~$u$ and~$v$ belong to the same component in the union-find data structure; (2)~$\dist(u,\vorbase(u)) \leq \cost(e_i)$; and (3)~$\dist(v,\vorbase(v)) \leq \cost(e_i)$.

This Voronoi-based test is very effective in random and dense Euclidean graphs, drastically reducing the graph size.

\subsection{Additional Results}

This section presents detailed results and data omitted from the main text due to space constraints.

Table~\ref{tab:sizes} presents more detailed information about each of the series tested in this work. Tables~\ref{tab:msserieserror} and \ref{tab:msseriestime} have results for our basic MS algorithms aggregated by series. Tables~\ref{tab:gmsserieserror} and \ref{tab:gmsseriestime} are similar, but refer to the GMS algorithm.

\begin{table}[htbp]\renewcommand{\arraystretch}{.85}\addtolength{\tabcolsep}{-2.1pt}
\small
\centering
\caption{Instance sizes. Minimum, rounded average, and maximum numbers of vertices, edges, and terminals for each series.}
\label{tab:sizes}
\begin{tabular}{llcrrrcrrrcrrr}
\toprule
\multicolumn{2}{c}{} & & \multicolumn{3}{c}{\tabhead{vertices}} & & \multicolumn{3}{c}{\tabhead{edges}} & & \multicolumn{3}{c}{\tabhead{terminals}} \\
\cmidrule(lr){4-6} \cmidrule(lr){8-10} \cmidrule(l){12-14}
\tabhead{class} & \tabhead{series} & & \tabhead{min}& \tabhead{avg}& \tabhead{max} & & \tabhead{min}& \tabhead{avg}& \tabhead{max} & & \tabhead{min}& \tabhead{avg}& \tabhead{max} \\
\midrule
\instance{euclidean} & \instance{p4e} & & 100 & 136 & 200 & & 4950 & 10386 & 19900 & & 5 & 26 & 100 \\
& \instance{p6e} & & 100 & 127 & 200 & & 180 & 231 & 370 & & 5 & 28 & 100 \\
& \instance{x} & & 52 & 259 & 666 & & 1326 & 74808 & 221445 & & 16 & 72 & 174\vspace{3 pt} \\
\instance{fst} & \instance{cph14} & & 16 & 2471 & 15473 & & 23 & 5969 & 38928 & & 10 & 143 & 1000 \\
& \instance{es0010} & & 12 & 17 & 24 & & 11 & 19 & 32 & & 10 & 10 & 10 \\
& \instance{es0020} & & 27 & 39 & 57 & & 26 & 47 & 83 & & 20 & 20 & 20 \\
& \instance{es0030} & & 43 & 69 & 118 & & 44 & 92 & 188 & & 30 & 30 & 30 \\
& \instance{es0040} & & 55 & 90 & 121 & & 55 & 121 & 180 & & 40 & 40 & 40 \\
& \instance{es0050} & & 83 & 116 & 143 & & 96 & 160 & 211 & & 50 & 50 & 50 \\
& \instance{es0060} & & 109 & 140 & 188 & & 133 & 192 & 280 & & 60 & 60 & 60 \\
& \instance{es0070} & & 142 & 167 & 209 & & 181 & 232 & 314 & & 70 & 70 & 70 \\
& \instance{es0080} & & 147 & 189 & 236 & & 180 & 259 & 343 & & 80 & 80 & 80 \\
& \instance{es0090} & & 175 & 217 & 284 & & 221 & 303 & 430 & & 90 & 90 & 90 \\
& \instance{es0100} & & 188 & 241 & 339 & & 233 & 336 & 522 & & 100 & 100 & 100 \\
& \instance{es0250} & & 542 & 624 & 713 & & 719 & 880 & 1053 & & 250 & 250 & 250 \\
& \instance{es0500} & & 1172 & 1303 & 1477 & & 1627 & 1871 & 2204 & & 500 & 500 & 500 \\
& \instance{es1000} & & 2532 & 2747 & 2984 & & 3615 & 4023 & 4484 & & 1000 & 1000 & 1000 \\
& \instance{es10000} & & 27019 & 27019 & 27019 & & 39407 & 39407 & 39407 & & 10000 & 10000 & 10000 \\
& \instance{tsp} & & 89 & 1756 & 17127 & & 104 & 2247 & 27352 & & 48 & 1130 & 11849\vspace{3 pt} \\
\instance{hard} & \instance{bip} & & 550 & 1690 & 3300 & & 3982 & 9013 & 18073 & & 50 & 190 & 300 \\
& \instance{cc} & & 64 & 1165 & 4096 & & 192 & 9797 & 28512 & & 8 & 112 & 473 \\
& \instance{hc} & & 64 & 1161 & 4096 & & 192 & 6418 & 24576 & & 32 & 581 & 2048 \\
& \instance{sp} & & 6 & 738 & 3997 & & 9 & 1974 & 10278 & & 3 & 408 & 2284\vspace{3 pt} \\
\instance{incidence} & \instance{i080} & & 80 & 80 & 80 & & 120 & 884 & 3160 & & 6 & 12 & 20 \\
& \instance{i160} & & 160 & 160 & 160 & & 240 & 3327 & 12720 & & 7 & 21 & 40 \\
& \instance{i320} & & 320 & 320 & 320 & & 480 & 12843 & 51040 & & 8 & 35 & 80 \\
& \instance{i640} & & 640 & 640 & 640 & & 960 & 50350 & 204480 & & 9 & 61 & 160\vspace{3 pt} \\
\instance{r} & \instance{1r} & & 1250 & 1250 & 1250 & & 2319 & 2341 & 2352 & & 6 & 32 & 60 \\
& \instance{2r} & & 2000 & 2000 & 2000 & & 5725 & 5766 & 5800 & & 9 & 51 & 98\vspace{3 pt} \\
\instance{random} & \instance{b} & & 50 & 75 & 100 & & 63 & 122 & 200 & & 9 & 23 & 50 \\
& \instance{c} & & 500 & 500 & 500 & & 625 & 4156 & 12500 & & 5 & 95 & 250 \\
& \instance{d} & & 1000 & 1000 & 1000 & & 1250 & 8312 & 25000 & & 5 & 186 & 500 \\
& \instance{e} & & 2500 & 2500 & 2500 & & 3125 & 20781 & 62500 & & 5 & 461 & 1250 \\
& \instance{mc} & & 97 & 261 & 400 & & 760 & 4208 & 11175 & & 45 & 126 & 213 \\
& \instance{p4z} & & 100 & 100 & 100 & & 4950 & 4950 & 4950 & & 5 & 18 & 50 \\
& \instance{p6z} & & 100 & 127 & 200 & & 180 & 231 & 370 & & 5 & 28 & 100\vspace{3 pt} \\
\instance{vienna} & \instance{gadv} & & 7565 & 27157 & 71184 & & 11521 & 42634 & 113616 & & 86 & 1918 & 6107 \\
& \instance{gori} & & 42481 & 109841 & 235686 & & 52552 & 157257 & 366093 & & 88 & 1979 & 6313 \\
& \instance{iadv} & & 160 & 12796 & 43690 & & 237 & 19045 & 66461 & & 23 & 1157 & 4138 \\
& \instance{isim} & & 1991 & 30129 & 89596 & & 3176 & 49017 & 148583 & & 38 & 1393 & 4991\vspace{3 pt} \\
\instance{vlsi} & \instance{alue} & & 940 & 8061 & 34479 & & 1474 & 12901 & 55494 & & 16 & 241 & 2344 \\
& \instance{alut} & & 387 & 10707 & 36711 & & 626 & 19741 & 68117 & & 34 & 161 & 879 \\
& \instance{diw} & & 212 & 3423 & 11821 & & 381 & 6434 & 22516 & & 10 & 20 & 50 \\
& \instance{dmxa} & & 169 & 1110 & 3983 & & 280 & 1959 & 7108 & & 10 & 15 & 23 \\
& \instance{gap} & & 179 & 1496 & 10393 & & 293 & 2579 & 18043 & & 10 & 22 & 104 \\
& \instance{lin} & & 53 & 8587 & 38418 & & 80 & 15956 & 71657 & & 4 & 31 & 172 \\
& \instance{msm} & & 90 & 1548 & 5181 & & 135 & 2682 & 8893 & & 10 & 17 & 89 \\
& \instance{taq} & & 122 & 2150 & 6836 & & 194 & 3648 & 11715 & & 10 & 39 & 136\vspace{3 pt} \\
\instance{wrp} & \instance{wrp3} & & 84 & 939 & 3168 & & 149 & 1833 & 6220 & & 11 & 51 & 99 \\
& \instance{wrp4} & & 110 & 571 & 1898 & & 188 & 1131 & 3616 & & 11 & 44 & 76 \\

\bottomrule
\end{tabular}
\end{table}

Table~\ref{tab:longsol} reports the best solution found considering all 13 runs (10 for MS2 and 3 for MSK) used to build Table~\ref{tab:longagg}. It includes the sizes of all instances tested in this experiment. Tables~\ref{tab:ms2sol} and~\ref{tab:ms2time} report results for all individual runs separately, for a more detailed view.

Tables~\ref{tab:i080}, \ref{tab:i160}, \ref{tab:i320}, \ref{tab:i640}, and \ref{tab:puc} show detailed results for our pure branch-and-bound algorithm. These runs do not use reduction-based preprocessing, which is generally ineffective for these instances.

\begin{table}[htbp]\renewcommand{\arraystretch}{.9}
\centering
\small
\caption{Multistart algorithm: Average percent error relative to best known solution.}
\label{tab:msserieserror}
\begin{tabular}{llrrrrrr}
\toprule
\tabhead{class} & \tabhead{series} & 1 & 4 & 16 & 64 & $256$ \\
\midrule
\instance{euclidean} & \instance{p4e} & 0.354079 & \optsol & \optsol & \optsol & $\optsol$ \\
& \instance{p6e} & 0.017396 & 0.016565 & \optsol & \optsol & $\optsol$ \\
& \instance{x} & 0.111395 & 0.002994 & 0.001633 & \optsol & $\optsol$\vspace{3 pt} \\
\instance{fst} & \instance{cph14} & 1.448691 & 0.742808 & 0.403645 & 0.173233 & $0.079506$ \\
& \instance{es0010} & 0.004975 & \optsol & \optsol & \optsol & $\optsol$ \\
& \instance{es0020} & 0.258192 & 0.001403 & 0.001403 & \optsol & $\optsol$ \\
& \instance{es0030} & 0.538273 & 0.095129 & \optsol & 0.007655 & $\optsol$ \\
\bottomrule
\end{tabular}
\end{table}

\begin{table}[htbp]\renewcommand{\arraystretch}{.9}
\centering
\small
\caption{Multistart algorithm: Average running times in seconds.}
\label{tab:msseriestime}
\begin{tabular}{llrrrrrr}
\toprule
\tabhead{class} & \tabhead{series} & 1 & 4 & 16 & 64 & $256$ \\
\midrule
\instance{euclidean} & \instance{p4e} & 0.006 & 0.014 & 0.047 & 0.169 & $0.633$ \\
& \instance{p6e} & 0.001 & 0.003 & 0.010 & 0.038 & $0.141$ \\
& \instance{x} & 0.016 & 0.041 & 0.127 & 0.496 & $1.999$\vspace{3 pt} \\
\instance{fst} & \instance{cph14} & 0.042 & 0.260 & 1.250 & 5.302 & $21.960$ \\
& \instance{es0010} & 0.000 & 0.001 & 0.002 & 0.005 & $0.020$ \\
& \instance{es0020} & 0.000 & 0.001 & 0.004 & 0.016 & $0.065$ \\
& \instance{es0030} & 0.001 & 0.003 & 0.010 & 0.040 & $0.169$ \\
\bottomrule
\end{tabular}
\end{table}

\begin{table}[p]\renewcommand{\arraystretch}{.9}
\centering
\small
\caption{Guarded Multistart algorithm: Average percent error relative to best known solution.}
\label{tab:gmsserieserror}
\begin{tabular}{llrrrrrr}
\toprule
\tabhead{class} & \tabhead{series} & 8 & 32 & 128 \\
\midrule
\instance{euclidean} & \instance{p4e} & \optsol & \optsol & \optsol \\
& \instance{p6e} & \optsol & \optsol & \optsol \\
& \instance{x} & \optsol & \optsol & \optsol\vspace{3 pt} \\
\instance{fst} & \instance{cph14} & 0.477741 & 0.256166 & 0.134948 \\
& \instance{es0010} & \optsol & \optsol & \optsol \\
& \instance{es0020} & \optsol & \optsol & \optsol \\
& \instance{es0030} & \optsol & \optsol & \optsol \\
\bottomrule
\end{tabular}
\end{table}

\begin{table}[p]\renewcommand{\arraystretch}{.9}
\centering
\small
\caption{Guarded Multistart algorithm: Average CPU times in seconds (wall times are lower).}
\label{tab:gmsseriestime}
\begin{tabular}{llrrrrrr}
\toprule
\tabhead{class} & \tabhead{series} & 8 & 32 & 128 \\
\midrule
\instance{euclidean} & \instance{p4e} & 0.009 & 0.009 & 0.009 \\
& \instance{p6e} & 0.004 & 0.005 & 0.004 \\
& \instance{x} & 0.045 & 0.045 & 0.045\vspace{3 pt} \\
\instance{fst} & \instance{cph14} & 1.139 & 5.056 & 20.791 \\
& \instance{es0010} & 0.001 & 0.001 & 0.002 \\
& \instance{es0020} & 0.002 & 0.002 & 0.002 \\
& \instance{es0030} & 0.006 & 0.005 & 0.004 \\
\bottomrule
\end{tabular}
\end{table}

\begin{table}[htbp]
\centering
\small
\caption{Best results found by long runs. The time~(in hours) is the \emph{sum} of the 13 executions of variants of our algorithm from Table~\ref{tab:longagg}. We also report the best previously known solution~(as of August 1, 2014) and the percent error. Negative errors favor our method.}
\label{tab:longsol}
\begin{tabular}{lrrr rrrr}
\toprule
\tabhead{name} & $|V|$ & $|E|$ & $|T|$ & \tabhead{time} [h] & \tabhead{known} & \tabhead{ours} & \tabhead{error} \\
\midrule
\instance{bip42p} & 1200 & 3982 & 200 & 12.74 & \textbf{24657} & 24764 & $0.434$ \\
\instance{bip42u} & 1200 & 3982 & 200 & 9.48 & \textbf{236} & \textbf{236} & $0.000$ \\
\instance{bip52p} & 2200 & 7997 & 200 & 28.05 & \textbf{24535} & 24628 & $0.379$ \\
\instance{bip52u} & 2200 & 7997 & 200 & 19.80 & \textbf{234} & 235 & $0.427$ \\
\instance{bip62p} & 1200 & 10002 & 200 & 25.14 & 22870 & \textbf{22843} & $-0.118$ \\
\instance{bip62u} & 1200 & 10002 & 200 & 22.11 & \textbf{220} & \textbf{220} & $0.000$ \\
\instance{bipa2p} & 3300 & 18073 & 300 & 67.19 & \textbf{35379} & 35413 & $0.096$ \\
\bottomrule
\end{tabular}
\end{table}

\begin{table}[htbp]
\centering
\small
\caption{Solutions found by MS2~(two-phase unguarded multistart algorithm) for open SteinLib instances, when varying the total number of iterations.}
\label{tab:ms2sol}
\begin{tabular}{lrrrrrr}
\toprule
\tabhead{name} & \tabhead{known} & \tabhead{1024} & \tabhead{4096} & \tabhead{16384} & \tabhead{65536} & \tabhead{262144} \\
\midrule
\instance{bip42p} & \textbf{24657} & 24888 & 24818 & 24811 & 24811 & 24811 \\
\instance{bip42u} & \textbf{236} & 237 & 237 & \textbf{236} & 237 & 237 \\
\instance{bip52p} & \textbf{24535} & 24775 & 24729 & 24771 & 24701 & 24628 \\
\instance{bip52u} & \textbf{234} & 237 & 235 & 235 & 235 & 235 \\
\instance{bip62p} & 22870 & 22924 & 22870 & \textbf{22843} & \textbf{22843} & \textbf{22843} \\
\instance{bip62u} & \textbf{220} & 221 & 221 & \textbf{220} & \textbf{220} & \textbf{220} \\
\instance{bipa2p} & \textbf{35379} & 35616 & 35555 & 35516 & 35523 & 35413 \\
\bottomrule
\end{tabular}
\end{table}

\begin{table}[htbp]
\centering
\small
\caption{Running time in seconds of MS2 (two-phase unguarded multistart algorithm) on open SteinLib instances, when varying the total number of iterations.}
\label{tab:ms2time}
\begin{tabular}{lrrrrr}
\toprule
\tabhead{name} & \tabhead{1024} & \tabhead{4096} & \tabhead{16384} & \tabhead{65536} & \tabhead{262144} \\
\midrule
\instance{bip42p} & 61 & 242 & 933 & 3790 & 15539 \\
\instance{bip42u} & 40 & 163 & 652 & 2696 & 11558 \\
\instance{bip52p} & 133 & 521 & 1951 & 7879 & 30967 \\
\instance{bip52u} & 80 & 325 & 1348 & 5551 & 22679 \\
\instance{bip62p} & 132 & 512 & 1943 & 8011 & 31593 \\
\instance{bip62u} & 96 & 384 & 1615 & 6692 & 27724 \\
\instance{bipa2p} & 329 & 1305 & 4850 & 18904 & 73150 \\
\bottomrule
\end{tabular}
\end{table}

\begin{table}[htbp]\renewcommand{\arraystretch}{.8}
\tiny
\begin{minipage}[t]{0.45\linewidth}\centering
\caption{Branch-and-bound on series \instance{i080}: dimensions, solution, branch-and-bound nodes, and running time in seconds.}
\label{tab:i080}
\begin{tabular}{lrrrrrr}
\toprule
\tabhead{instance} & $|V|$ & $|E|$ & $|T|$ & \tabhead{opt} & \tabhead{bb} & \tabhead{time} [s] \\
\midrule
\instance{i080-001} & 80 & 120 & 6 & 1787 & 1 & 0.000 \\
\instance{i080-002} & 80 & 120 & 6 & 1607 & 2 & 0.001 \\
\instance{i080-003} & 80 & 120 & 6 & 1713 & 1 & 0.000 \\
\instance{i080-004} & 80 & 120 & 6 & 1866 & 5 & 0.001 \\
\instance{i080-005} & 80 & 120 & 6 & 1790 & 2 & 0.000 \\
\instance{i080-011} & 80 & 350 & 6 & 1479 & 3 & 0.001 \\
\instance{i080-012} & 80 & 350 & 6 & 1484 & 4 & 0.001 \\
\bottomrule
\end{tabular}
\end{minipage}\hspace{0.1\linewidth}
\begin{minipage}[t]{0.45\linewidth}\centering
\caption{Branch-and-bound on series \instance{i160}: dimensions, solution, branch-and-bound nodes, and running time in seconds.}
\label{tab:i160}
\begin{tabular}{lrrrrrr}
\toprule
\tabhead{instance} & $|V|$ & $|E|$ & $|T|$ & \tabhead{opt} & \tabhead{bb} & \tabhead{time} [s] \\
\midrule
\instance{i160-001} & 160 & 240 & 7 & 2490 & 5 & 0.002 \\
\instance{i160-002} & 160 & 240 & 7 & 2158 & 1 & 0.000 \\
\instance{i160-003} & 160 & 240 & 7 & 2297 & 1 & 0.001 \\
\instance{i160-004} & 160 & 240 & 7 & 2370 & 1 & 0.001 \\
\instance{i160-005} & 160 & 240 & 7 & 2495 & 2 & 0.001 \\
\instance{i160-011} & 160 & 812 & 7 & 1677 & 5 & 0.003 \\
\instance{i160-012} & 160 & 812 & 7 & 1750 & 5 & 0.003 \\
\bottomrule
\end{tabular}
\end{minipage}
\end{table}

\begin{table}[htbp]\renewcommand{\arraystretch}{.8}\addtolength{\tabcolsep}{-1pt}
\tiny
\begin{minipage}[t]{0.45\linewidth}\centering
\caption{Branch-and-bound on series \instance{i320}: dimensions, solution, branch-and-bound nodes, and running time in seconds.}
\label{tab:i320}
\begin{tabular}{lrrrrrr}
\toprule
\tabhead{instance} & $|V|$ & $|E|$ & $|T|$ & \tabhead{opt} & \tabhead{bb} & \tabhead{time} [s] \\
\midrule
\instance{i320-001} & 320 & 480 & 8 & 2672 & 1 & 0.001 \\
\instance{i320-002} & 320 & 480 & 8 & 2847 & 1 & 0.001 \\
\instance{i320-003} & 320 & 480 & 8 & 2972 & 5 & 0.002 \\
\instance{i320-004} & 320 & 480 & 8 & 2905 & 5 & 0.002 \\
\instance{i320-005} & 320 & 480 & 8 & 2991 & 5 & 0.004 \\
\instance{i320-011} & 320 & 1845 & 8 & 2053 & 16 & 0.013 \\
\instance{i320-012} & 320 & 1845 & 8 & 1997 & 5 & 0.007 \\
\bottomrule
\end{tabular}
\end{minipage}\hspace{0.1\linewidth}
\begin{minipage}[t]{0.45\linewidth}\centering
\caption{Branch-and-bound on series \instance{i640} (excluding \instance{i640-31*}): dimensions, solution, branch-and-bound nodes, and running time in seconds.}
\label{tab:i640}
\begin{tabular}{lrrrrrr}
\toprule
\tabhead{instance} & $|V|$ & $|E|$ & $|T|$ & \tabhead{opt} & \tabhead{bb} & \tabhead{time} [s] \\
\midrule
\instance{i640-001} & 640 & 960 & 9 & 4033 & 2 & 0.002 \\
\instance{i640-002} & 640 & 960 & 9 & 3588 & 3 & 0.004 \\
\instance{i640-003} & 640 & 960 & 9 & 3438 & 2 & 0.002 \\
\instance{i640-004} & 640 & 960 & 9 & 4000 & 5 & 0.009 \\
\instance{i640-005} & 640 & 960 & 9 & 4006 & 4 & 0.006 \\
\instance{i640-011} & 640 & 4135 & 9 & 2392 & 1 & 0.011 \\
\instance{i640-012} & 640 & 4135 & 9 & 2465 & 8 & 0.014 \\
\bottomrule
\end{tabular}
\end{minipage}
\end{table}

\begin{table}[p]
\centering
\caption{Branch-and-bound results on select \instance{hard} instances: dimensions, solution, branch-and-bound nodes, and running time in seconds.}
\label{tab:puc}
\begin{tabular}{lrrrrrr}
\toprule
\tabhead{instance} & $|V|$ & $|E|$ & $|T|$ & \tabhead{opt} & \tabhead{bb} & \tabhead{time} [s] \\
\midrule
\instance{bipe2p} & 550 & 5013 & 50 & 5616 & 158071 & 202.060 \\
\instance{bipe2u} & 550 & 5013 & 50 & 54 & 473584 & 248.572 \\
\instance{cc3-4p} & 64 & 288 & 8 & 2338 & 339 & 0.040 \\
\instance{cc3-4u} & 64 & 288 & 8 & 23 & 273 & 0.018 \\
\instance{cc3-5p} & 125 & 750 & 13 & 3661 & 20660 & 4.399 \\
\instance{cc3-5u} & 125 & 750 & 13 & 36 & 39892 & 3.669 \\
\instance{cc5-3p} & 243 & 1215 & 27 & 7299 & 2102429 & 1196.743 \\
\instance{cc5-3u} & 243 & 1215 & 27 & 71 & 1904887 & 512.745 \\
\instance{cc6-2p} & 64 & 192 & 12 & 3271 & 213 & 0.025 \\
\instance{cc6-2u} & 64 & 192 & 12 & 32 & 167 & 0.012 \\
\instance{hc6p} & 64 & 192 & 32 & 4003 & 3036 & 0.250 \\
\instance{hc6u} & 64 & 192 & 32 & 39 & 3215 & 0.194 \\
\instance{hc7p} & 128 & 448 & 64 & 7905 & 1979435 & 366.055 \\
\instance{hc7u} & 128 & 448 & 64 & 77 & 6471809 & 768.549 \\

\bottomrule
\end{tabular}
\end{table}

\end{document}